\documentclass[12pt]{article}
\usepackage{graphicx}

\newcommand\pubnumber{DPF2015-394}
\newcommand\pubdate{\today}

\def\umn{School of Physics and Astronamy\\
University of Minnesota, MN 55455, USA}

\def\Title#1{\begin{center} {\Large #1 } \end{center}}
\def\Author#1{\begin{center}{ \sc #1} \end{center}}
\def\Address#1{\begin{center}{ \it #1} \end{center}}

\newcommand\pubblock{\rightline{\begin{tabular}{l} \pubnumber\\
         \pubdate  \end{tabular}}}
\newenvironment{Abstract}{\begin{quotation}  }{\end{quotation}}
\newenvironment{Presented}{\begin{quotation} \begin{center} 
             PRESENTED AT\end{center}\bigskip 
      \begin{center}\begin{large}}{\end{large}\end{center} \end{quotation}}

\def\beq{\begin{equation}}
\def\eeq#1{\label{#1}\end{equation}}
\def\eeqn{\end{equation}}

\def\beqa{\begin{eqnarray}}
\def\eeqa#1{\label{#1}\end{eqnarray}}
\def\eeqan{\end{eqnarray}}

\let\bar=\overbar

\def\Dslash{\not{\hbox{\kern-4pt $D$}}}
\def\dslash{\not{\hbox{\kern-2pt $\del$}}}

\def\msb{{\bar{\ssstyle M \kern -1pt S}}}

\begin{document}
\begin{titlepage}
\pubblock

\vfill
\Title{The CAPTAIN Experiment}
\vfill
\Author{ Jianming Bian}
\Address{\umn}
\vfill
\begin{Abstract}
The Cryogenic Apparatus for Precision Tests of Argon Interactions
with Neutrinos (CAPTAIN) program is designed to make measurements
of scientific importance to long-baseline neutrino physics and physics top-
ics that will be explored by large underground detectors. The experiment
employs two liquid Argon time projection chambers (LArTPCs), a pri-
mary detector with a mass of approximately 10 ton that will be deployed
at different facilities for physics measurements and a two tonprototype
detector for configuration testing. The physics programs for CAPTAIN
include measuring neutron interactions at the Los Alamos Neutron Sci-
ence Center, measuring neutrino interactions in the high-energy regime
(1.5$-$5 GeV) at FermilabÕs NuMI beam, and measuring neutrino interac-
tions in the low-energy regime ($<$50 MeV) at stopped pion sources for
supernova neutrino studies. The prototype detector (Mini-CAPTAIN) has
been commissioned and the first UV laser track has been seen in its TPC.
This paper gives an overview of the CAPTAIN program and reports the
status of the commissioning. The up-to-date detector design and running
plans are also described.
\end{Abstract}
\vfill
\begin{Presented}
DPF 2015\\
The Meeting of the American Physical Society\\
Division of Particles and Fields\\
Ann Arbor, Michigan, August 4--8, 2015\\
\end{Presented}
\vfill
\end{titlepage}
\def\thefootnote{\fnsymbol{footnote}}
\setcounter{footnote}{0}

\section{Introduction}

The CAPTAIN (Cryogenic Apparatus for Precision Tests of Argon Interactions with Neutrino) experiment~\cite{Berns:2013usa}  is designed to make measurements of scientific importance to long-baseline neutrino physics and other physics topics that will be explored by large underground detectors. It began as part of a Los Alamos National Laboratory (LANL) Laboratory Directed Research and Development (LDRD) project and has evolved into a multi-institutional collaboration. This program consists of two detectors, CAPTAIN and Mini-CAPTAIN. The CAPTAIN detector is a Liquid Argon Time Project Chamber (LArTPC) deployed in a portable and evacuable cryostat that can hold a total of 7700 liters of liquid argon. The Mini-CAPTAIN detector is a prototype of the CAPTAIN detector that contains 1700 liters of liquid argon. The CAPTAIN project plans to study interactions in liquid argon with a neutron source at LANL and neutrino beams at Fermi National Accelerator Laboratory (Fermilab). These measurements can significantly benefit the development of the Deep Underground Neutrino Experiment (DUNE)~\cite{Adams:2013qkq} . In particular, the CAPTAIN program impacts several of the topics that make up the DUNE physics program via two prongs of study: low-energy neutrino physics and high-energy neutrino physics.

\section{Detectors}

\begin{figure}[htb]
\centering
\includegraphics[height=3.5in]{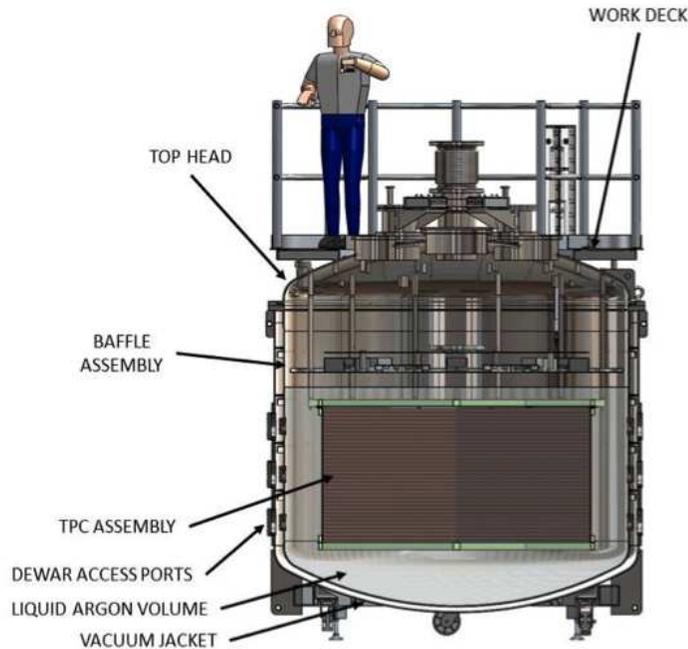}
\caption{Schematic drawing of the CAPTAIN detector}
\label{fig:CAPTAINDetector}
\end{figure}

The CAPTAIN detector and the Mini-CAPTAIN detector are similar, including Cryostats, Cryogenics, Electronics, TPCs, Photon Detection System and Laser Calibration System. The design differences between the two detectors are driven by the cryostat sizes and geometries. The Mini-CAPTAIN prototype uses a 1700-liter vacuum-jacketed cryostat provided by UCLA. The vacuum jacket is 60.25 inches in diameter, and the vessel is 64.4 inches in height. The CAPTAIN detector uses a 7700L vacuum insulated liquid argon cryostat that is 107.5 inches in diameter and 115 inches in height. The schematic of the CAPTAIN cryostat and TPC are shown in Figure~\ref{fig:CAPTAINDetector}. The designed maximum electron drift distance is 100.0 cm for the full CAPTAIN detector and 32.0 cm for Mini-CAPTAIN. Accordingly, the equivalent O2 contamination is required to be smaller than 750 ppt for mini-CAPTAIN and 240 ppt for the CAPTAIN detecter.

%%%%%%%%%%%%%%%%%%%%%%%%%%%%%%%%%%%%%%%%%%%%%%%%%%%%%%%%%%%%%%%%%%%%%%%%%
%%
%%   use this format to include an .pdf figure into your paper
%%
%%%%%%%%%%%%%%%%%%%%%%%%%%%%%%%%%%%%%%%%%%%%%%%%%%%%%%%%%%%%%%%%%%%%%%%%%%%

\begin{figure}[htb]
\centering
\includegraphics[height=3.8in]{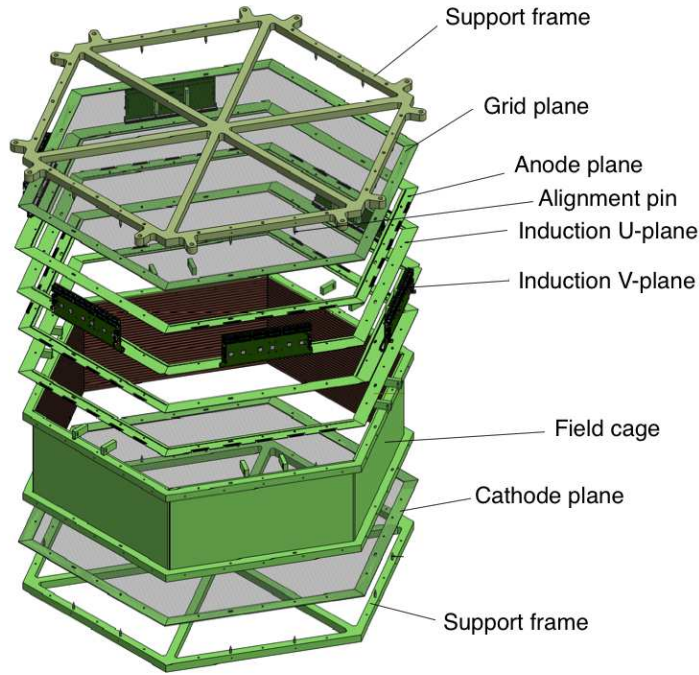}
\caption{Component detailed view of the CAPTAIN TPC}
\label{fig:TPCfigure}
\end{figure}

As shown in Figure \ref{fig:TPCfigure}, the CAPTAIN TPC consists of a field cage in a hexagonal shape with a mesh cathode plane on the bottom of the hexagon and a series of four wire planes (grid, U, V and anode planes) on the top with a mesh ground plane. The apothem of the TPC is 100 cm and the drift length between the anode and cathode is 100 cm. Each wire plane has 667 copper beryllium wires that are 75 $\mu$m in diameter. The space between two wires is 3 mm within a plane and is 3.125 mm between two planes. The U and V wires are oriented at $\pm$60 degrees with respect to the anode wires, allowing two dimensional detection of electrons passing through by charge induction. The third coordinate is determined by the drift time to the anode plane. The design voltage gradient provided by the field cage is 500 V/cm when 50 kV is applied to the cathode. Under these conditions the drift velocity of the electrons is 1.6 mm/$\mu$s. The TPC in Mini-CAPTAIN is a smaller version of the CAPTAIN TPC. The drift length of Mini-CAPTAIN TPC is 32 cm and the apothem is 50 cm. Each wire plane has 337 wires.

The electronic components for the TPC are identical to those of the MicroBooNE experiment at Fermilab~\cite{ref:microboone}. The front-end mother board is designed with twelve custom CMOS Application Specific Integrated Circuits (ASIC). Each ASIC reads out 16 channels from the TPC. The mother board is mounted directly on the TPC wire planes and is designed to be operated in liquid argon. The output signals from the mother board are transmitted through the cold cables to the cryostat feed-thru to the intermediate amplifier board. The intermediate amplifier is designed to drive the differential signals through long cable lengths to the 64 channel receiver ADC board. The digital signal is then processed in an FPGA on the Front End Module (FEM) board. All signals are transmitted via fiber optic cable from a transmit module to the data acquisition computer.

The CAPTAIN and Mini-CAPTAIN detectors are both equipped with photon detection systems. These improve the energy measurement by taking advantage of the anti-correlation between the scintillation photons detected by the PMT and the ionization electrons detected by the TPC. This system uses tetraphenyl butadiene (TPB) as a wavelength shifter to convert the 128 nm scintillation photons in liquid argon to a re-emission photon spectrum that peaks at about 420 nm. Hamamatsu R8520-500 photomultiplier tubes (PMT) are used to detect the wave-shifted light. The R8520 is a compact PMT approximately 1''$\times$1'' $\times$ 1'' in size with a borosilicate glass window and a special bialkali photocathode capable of operation at liquid argon temperatures (87 K). The TPB is coated on a thin piece of acrylic in front of the PMTs. Sixteen PMTs are deployed for both CAPTAIN and Mini-CAPTAIN detectors, 8 on top of the TPC volume and 8 on the bottom. Light yield is greater than 2.2 photoelectrons per MeV for a minimum ionizing particle (MIP).

\begin{figure}[htb]
\centering
\includegraphics[height=2.7in]{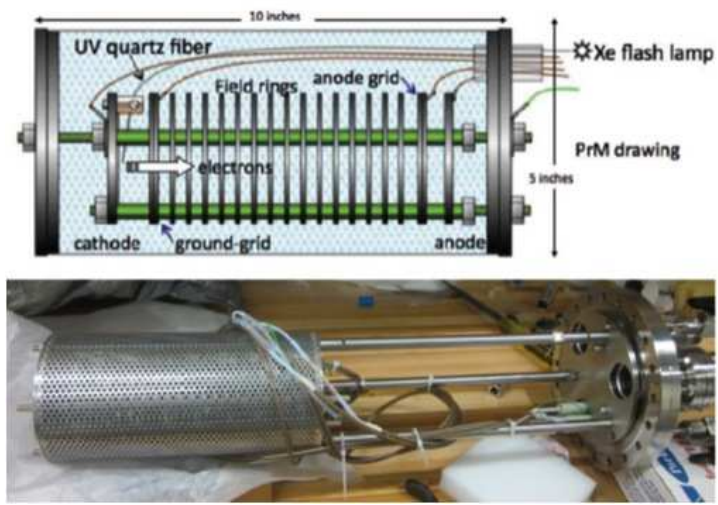}
\caption{Drawing (top) and assembly (bottom) of the purity monitor}
\label{fig:prm}
\end{figure}

Purification is essential for successful operation of liquid argon TPC. CAPTAIN uses a purity monitor built by University of Minnesota to continuously monitor the purity of liquid argon during the running. The design of the purity monitor is based on these used at the Fermilab Liquid Argon Purity Demonstrator (LAPD)~\cite{Adamowski:2014daa}. The drawing and assembly of the purity monitor are shown in Figure~\ref{fig:prm}. It consists of four parallel, circular electrodes: a disk holding a photocathode, two wire grid rings for anode and cathode, and an anode disk. The region between the anode grid and cathode grid contains a series of field-shaping stainless steel rings that are electrically connected by resistors, providing a voltage gradient of 100 V/cm when 2500V is applied to the cathode. A stainless mesh cylinder is used as a Faraday cage to isolate the purity monitor from electrostatic backgrounds. The electronics includes a filter circuit and an amplifier circuit that are separated by a copper plate and two D0 calorimeter amplifiers are used to amplify signals. The light system includes a Hamamatsu Xenon flash lamp and a UV quartz fiber that carries the light to illuminate the gold photocathode. The electron drift lifetime, which is inversely proportional to purity, can be determined from the ratio of the cathode signal to the anode signal and the time difference between the two signals. During operation, the purity monitor is mounted on the side of the TPC.

A existing LANL Quantel Brilliant B Q-switched Nd:YAG laser is used to ionize the liquid Argon. It creates well-defined ionization tracks within the liquid Argon TPC that are an excellent calibration source for electron lifetime measurements in TPC. The original wavelength and pulse energy are 1064 nm and 850 mJ, respectively. Using frequency doublers, the wavelength of the laser will be changed to 266 nm (pulse energy 90 mJ) to produce ionization tracks in the liquid argon.

\section{Physics and Running Plan}

\subsection{Neutron running at LANL}
The CAPTAIN program plans to perform neutron studies at the Weapons Neutron Research Facility (WNR) at LANL. The project plans to place the Mini-CAPTAIN detector in the WNR neutron beam and will characterize neutron interactions in argon. The running with WNR neutron beam includes both high-energy neutrons and low-energy neutrons, which will help DUNE to understand (1) high-energy neutrino and neutron detection for neutrino oscillation studies at DUNE (2) low-energy neutrino and neutron detection, for supernova neutrino studies at DUNE.

At a distance of 1300 km, the DUNE experiment will measure neutrino interactions in the energy range 1.5$-$5 GeV, close to the first oscillation maximum. In this energy region, neutrino-nucleus interactions are not well understood. Previously, the ArgoNEUT experiment made the first and still only inclusive measurement with the NuMI beam. For the exclusive interaction modes, rich and complex neutrino-nuclei interactions are dominated (more than half) by the baryon resonance channel. One of the hardest parts of understanding these interactions is the behavior of neutrons produced in these final states. In the WNR neutron beam, Mini-CAPTAIN can study the high-energy neutrons that produce pions, helping to develop techniques to identify neutron interactions and reconstruct neutrino energy in liquid argon detectors.

Precise measurement of supernova neutrinos is also an important physics topic for the DUNE experiment. The measurement of the time evolution of the energy and flavor spectrum of neutrinos from supernovae can revolutionize our understanding of neutrino properties, supernova physics, and possibly discover or tightly constrain non-standard neutrino interactions. The detection of a galactic supernova neutrino burst in DUNE requires the capability to tag and identify the following charged-current (CC) and neutral-current (NC) interactions. The low-energy neutron interactions collected by Mini-CAPTAIN at WNR can be used to study the neutrino-like argon reaction: $n + ^{40}Ar\to^{40}Ar^* + n$. This interaction is a very good control sample for the neutral-current (NC) interactions,  $\nu_x + ^{40}Ar\to^{40}Ar^* + \nu_x$, introduced by supernova neutrinos.

\subsection{CAPTAIN-MINERvA in the NuMI beam}

The NuMI neutrino beam at Fermilab was upgraded for the NOvA, MINERvA and MINOS experiments. On-axis NuMI provides a broad neutrino energy spectrum from 1 to 10 GeV for MINERvA and MINOS. Off-axis it provides a narrow peak from 1 to 3 GeV for the NOvA experiment. The NuMI running of the CAPTAIN detector in both a neutrino and antineutrino beam is an integral part of understanding the neutrino cross sections needed by DUNE, and liquid argon detectors in general, to interpret neutrino oscillation signals. The CAPTAIN detector will be positioned right in front of the MINERvA detector where it will measure on-axis NuMI beam neutrinos. With this setup, the MINERvA detector (+ MINOS Near Detector) will act as a calorimeter for the CAPTAIN detector and measure the energy of the final state particles that exit the CAPTAIN detector. Since the MicroBooNE experiment detects neutrinos below 1 GeV that are produced by the Booster Neutrino Beam (BNB) beam, the combination of CAPTAIN and MINERvA in the higher-energy NuMI beam can provide complementary measurements to MicroBooNE for the full DUNE energy spectrum. CAPTAIN-MINERvA running will also improvement measurements of flux/cross-section that benefit existed NuMI experiments like NOvA. The CAPTAIN-MINERvA proposal was presented at the last Fermilab PAC meeting and the PACs recommendation to proceed has been accepted by the Fermilab director.

\subsection{Low-energy neutrino study with the stopped pion source at the BNB}

The Booster Neutrino Beam (BNB) at FNAL was designed and built as a conventional neutrino beam with a decay region to produce pion-decay-in-fight neutrinos for the MiniBooNE experiment, and will also run to support the MicroBooNE experiment. Due to the short decay region, it also can serve as as a source of neutrinos from stopped pions in the target, horn and surrounding structures. We plan to put the CAPTAIN detector close enough to collect low-energy neutrinos produced by these pions that decay at rest. This will be the first time that a liquid argon TPC is run for low-energy neutrinos, and it will be a great help for DUNE supernova study. It is possible that the CAPTAIN detector at the BNB could be put as close as 10 m to the absorber and we expect to have charged current (CC) neutrino absorption event rates of about 200 per year for $2\times10^20$ POT. However, one challenge posed by operating CAPTAIN so close to the absorber are the neutron backgrounds, and we plan to use the SciBath detector to determine whether we need to provide shielding to mitigate them. In addition, we could triple the event rate with lower neutron backgrounds if the beam is run in the off-target mode.

\section{Mini-CAPTAIN commissioning}

\begin{figure}[htb]
\centering
\includegraphics[height=2.6in]{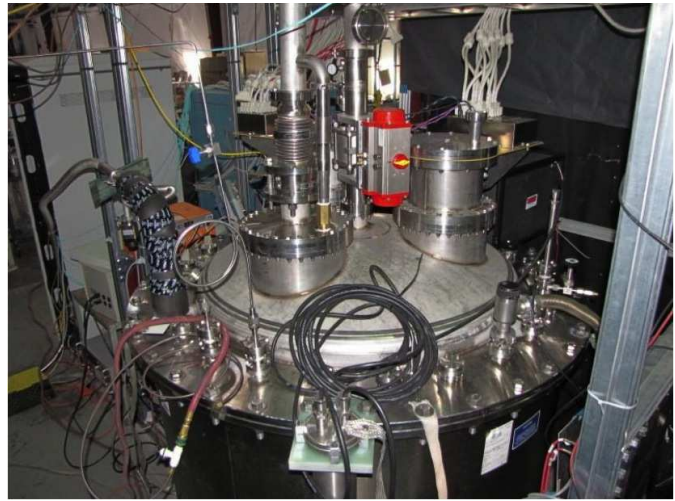}
\caption{Mini-CAPTAIN Assembly}
\label{fig:minicaptainassembly}
\end{figure}

\begin{figure}[htb]
\centering
\includegraphics[height=2.1in]{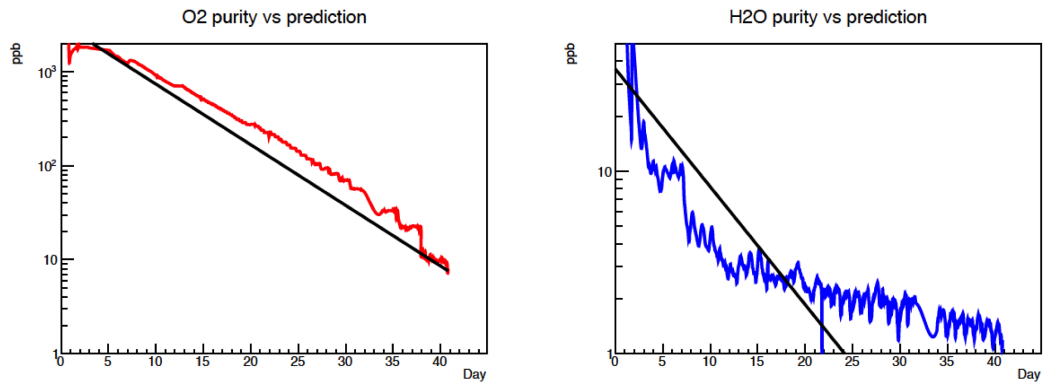}
\caption{Purity vs. time for Mini-CAPTAIN}
\label{fig:purity}
\end{figure}

The cryostat, cryogenics and TPC of the Mini-CAPTAIN detector have been commissioned as shown in Figure~\ref{fig:minicaptainassembly}. The electronics, TPC and heat load have been tested. The photon detector system and purity monitor will be installed soon. Purification tests for Mini-CAPTAIN have been under way since July 2015. Because the purity monitor has not been installed, we use a DF 310E Process Oxygen Analyzer (Servomex Company Inc.) as the O$_2$ monitor and a Continuous Wave CRDS Trace Gas Analyzer (Tiger Optics, LLC.) as the H2O monitor. Sensitivities for O$_2$ and H$_2$O are 3 ppb and 1 ppb, respectively. The Mini-CAPTAIN cryogenics and purification system consists of an inline liquid argon filter and a gas recirculation system including a condenser and filters. During the filling procedure, industrial high-purity liquid argon (O$_2$ 2.7 ppm, H$_2$O 0.5 ppm) is passed through the inline filter where the molecular sieve removes water and the copper removes O$_2$. As liquid argon is boiled off in the ullage volume, the argon vapor is carried by a 1/2 inch line to the gas recirculation system. We use the two SAES argon gas purifiers, model MC1500, configured in parallel to purify the boil-off argon gas, then a liquid nitrogen condenser liquifies the purified argon gas for recycling back to the liquid argon volume. Figure~\ref{fig:purity} shows measured purities (colored line) and theoretically calculated purities (black line) measured by the O$_2$ and H$_2$O analyzer until Aug. 1, 2015, 40 days after the filling of liquid argon. A Mini-CAPTAIN neutron run is scheduled for January of 2016.

\begin{figure}[htb]
\centering
\includegraphics[height=2.0in]{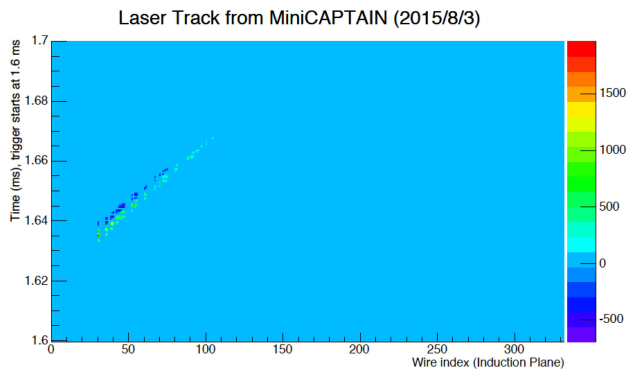}
\caption{First laser track observed by the Mini-CAPTAIN TPC on Aug. 3, 2015}
\label{fig:lasertrack}
\end{figure}

On Aug. 3, 2015, the first ionization track from the laser calibration system was observed by the Mini-CAPTAIN TPC. The first image of a UV laser track recorded by a induction plane of the TPC is shown in Figure~\ref{fig:lasertrack}. The observation of the first track demonstrates that the purity has achieved the work region of the liquid argon TPC. The overall purity read by the analyzers is 5 ppb.

\section{Status of the CAPTAIN detector}
\begin{figure}[htb]
\centering
\includegraphics[height=2.0in]{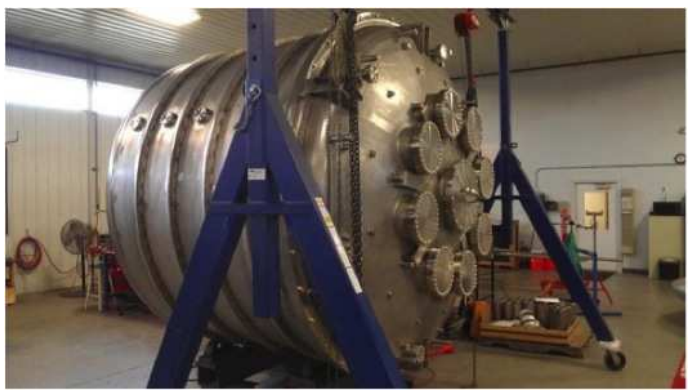}
\caption{The cryostat of the CAPTAIN detector}
\label{fig:captaincryo}
\end{figure}

\begin{figure}[htb]
\centering
\includegraphics[height=2.0in]{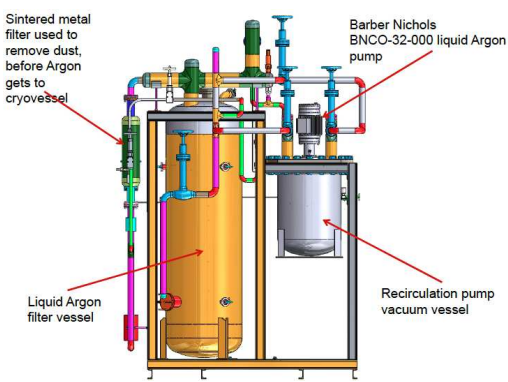}
\caption{The design of purification system for CAPTAIN detector}
\label{fig:purification}
\end{figure}

The cryostat, electronics and field cage are in hand to construct the CAPTAIN detector. Preliminary checks of their quality have been done. Figure~\ref{fig:captaincryo} shows the cryostat of CAPTAIN that has been delivered to LANL. Unlike the purification system that circulates and filters evaporated argon gas in the Mini-CAPTAIN detector, CAPTAIN will use a liquid cryogenic pump to circulate liquid argon and filter the liquid argon directly. In this way the speed of the purification process will not be limited by the evaporation and re-condensation. The CAPTAIN purification system, as shown in Figure~\ref{fig:purification}, includes a filter vessel, a pumping system, a dust removal filter, and the skid. We chose Barber Nichols BNCP-32B-000 liquid argon centrifugal pump, which has a pump capacity of about 10 gal/min as the liquid cryogenic pump, and a recirculation pump vacuum vessel is attached to the pump. A sintered metal filter from Purolator Facet Inc. is used to remove the dust that comes from the purification system. Two filter volumes are designed in the filter vessel: one for removal of oxygen using BASF CU-0226 copper oxide impregnated spheres, 14-28 mesh; and one for removal of water using Aldrich Molecular Sieve type 4A. The major components of the CAPTAIN purification system described above are at the vendor and are expected to be delivered to LANL in fall 2015. The earliest date that CAPTAIN could be moved to Fermilab would be Fall 2016.

\section{Summary}

The CAPTAIN program will make significant contributions to the development of capabilities for future DUNE physics. It will investigate neutrino and neutron interactions in the low-energy region ($<$ 50 MeV) and high-energy region (1.5$-$5 GeV). These measurements with CAPTAIN andMini-CAPTAIN detectors will provide great benefit for the neutrino oscillation and supernova neutrino studies at DUNE. Commissioning of the Mini-CAPTAIN detector is reaching completion and its neutron beam running will take place in January 2016. A full proposal to the Fermilab PAC for the CAPTAIN-MINERvA project has been submitted and the Stage 1 approval from the Fermilab director has been received.

\end{document}